\documentclass[sigconf]{acmart}

\AtBeginDocument{%
  \providecommand\BibTeX{{%
    \normalfont B\kern-0.5em{\scshape i\kern-0.25em b}\kern-0.8em\TeX}}}

\usepackage{todonotes}
\usepackage[T1]{fontenc}
\usepackage{caption}
\usepackage{subcaption}

\usepackage{tikz}
\usetikzlibrary{spy}
\usepackage{pgfplots}
\usetikzlibrary{shapes,arrows}
\tikzstyle{line}=[draw]

\copyrightyear{2021}
\acmYear{2021}
\setcopyright{acmcopyright}\acmConference[LSC '21]{Proceedings of the 4th Annual Lifelog Search Challenge}{August 21, 2021}{Taipei, Taiwan}
\acmBooktitle{Proceedings of the 4th Annual Lifelog Search Challenge (LSC '21), August 21, 2021, Taipei, Taiwan}
\acmPrice{15.00}
\acmDOI{10.1145/3463948.3469060}
\acmISBN{978-1-4503-8533-6/21/08}

\begin{document}

\title{lifeXplore at the Lifelog Search Challenge 2021}
\renewcommand{\shorttitle}{lifeXplore at the Lifelog Search Challenge 2021}

\author{Andreas Leibetseder, \mbox{Klaus Schoeffmann}}
\affiliation{%
  \institution{Institute of Information Technology}
  \institution{Alpen-Adria University, 9020 Klagenfurt Austria}
}
\email{{aleibets|ks}@itec.aau.at}

\renewcommand{\shortauthors}{Leibetseder et al.}


%
%




\fancyhead{}

\begin{abstract}

Since its first iteration in 2018, the Lifelog Search Challenge (LSC) continues to rise in popularity as an interactive lifelog data retrieval competition, co-located at the ACM International Conference on Multimedia Retrieval (ICMR). The goal of this annual live event is to search a large corpus of lifelogging data for specifically announced memories using a purposefully developed tool within a limited amount of time. As long-standing participants, we present our improved lifeXplore -- a retrieval system combining chronologic day summary browsing with interactive combinable concept filtering. Compared to previous versions, the tool is improved by incorporating temporal queries, advanced day summary features as well as usability improvements.

\end{abstract}

\maketitle

\section{Introduction} 









Personal lifelogs are periodically recorded data collections that document a person's life at a particular level of detail. Typically they include images captured from a first-person point of view but potentially as well other data such as a geographical location, an individual's vital signs or even explicitly added dietary information. Expectedly, the daily process of maintaining such a thorough documentation not only requires consistency and careful planning, it also heavily depends on reliable, sufficient and well-organized data storage. Albeit given an appropriate strategy, filing data on a daily basis appears to be a manageable task, however, revisiting that data in a meaningful way can be arbitrarily complex, especially for archives containing decades of information. This is due to the observation that, generally, the human brain is very effective in storing memories but recalling them often is not as precise as we'd hope it to be: apart from remembering events in our lives with varying degrees of accuracy, we especially often lack to recall the exact point in time they have occurred, i.e. the dates and/or time of day. This stands in contrast to the way we tend to store data on our computers: for instance, a typical personal lifelog is saved on a daily basis and usually organized by that date -- quickly retrieving moments would imply knowing their particular timeframes or dates. While this may be possible in some cases, it is much too restrictive and renders searching lifelog data time-consuming and tedious.

The annual Lifelog Search Challenge (LSC)~\cite{gurrin2019comparing,LSC20,LSC21} addresses these issues by providing international teams of researches with a platform for developing strategies aiding the improvement of lifelog data collection search and retrieval. Inspired by another long-standing popular multimedia retrieval competition, the Video Browser Showdown (VBS)~\cite{VBS2014,Lokoc2018,lokoc2019interactive}, the LSC is held as a live event, where participating teams are tasked with solving several time-constrained search assignments using their custom-developed systems on a lifelogging database~\cite{gurrin2019test,LSC20} that spans approximately 4 months of anonymized data. This data is available to the participants during the development phase, hence, it can be used for pre-processing according to the individual teams' preferences and requirements. Similarly to the previous LSC iteration (LSC2020), the current dataset~\cite{LSC21} comprises periodically (approx. every 40 seconds) recorded images from the lifeloggers' perspective together with metadata such as location, heart rate and automatically extracted deep visual concepts. Finally, since the systems are encouraged to be developed keeping usability in mind, not only the developers partake at the LSC as \textit{experts} but also voluntary \textit{novice} users that are unfamiliar with the systems compete against each other. 

\begin{figure*}[htbp!]

    \centering
    \begin{subfigure}{0.5\textwidth}
      \centering
      \includegraphics[width=\linewidth]{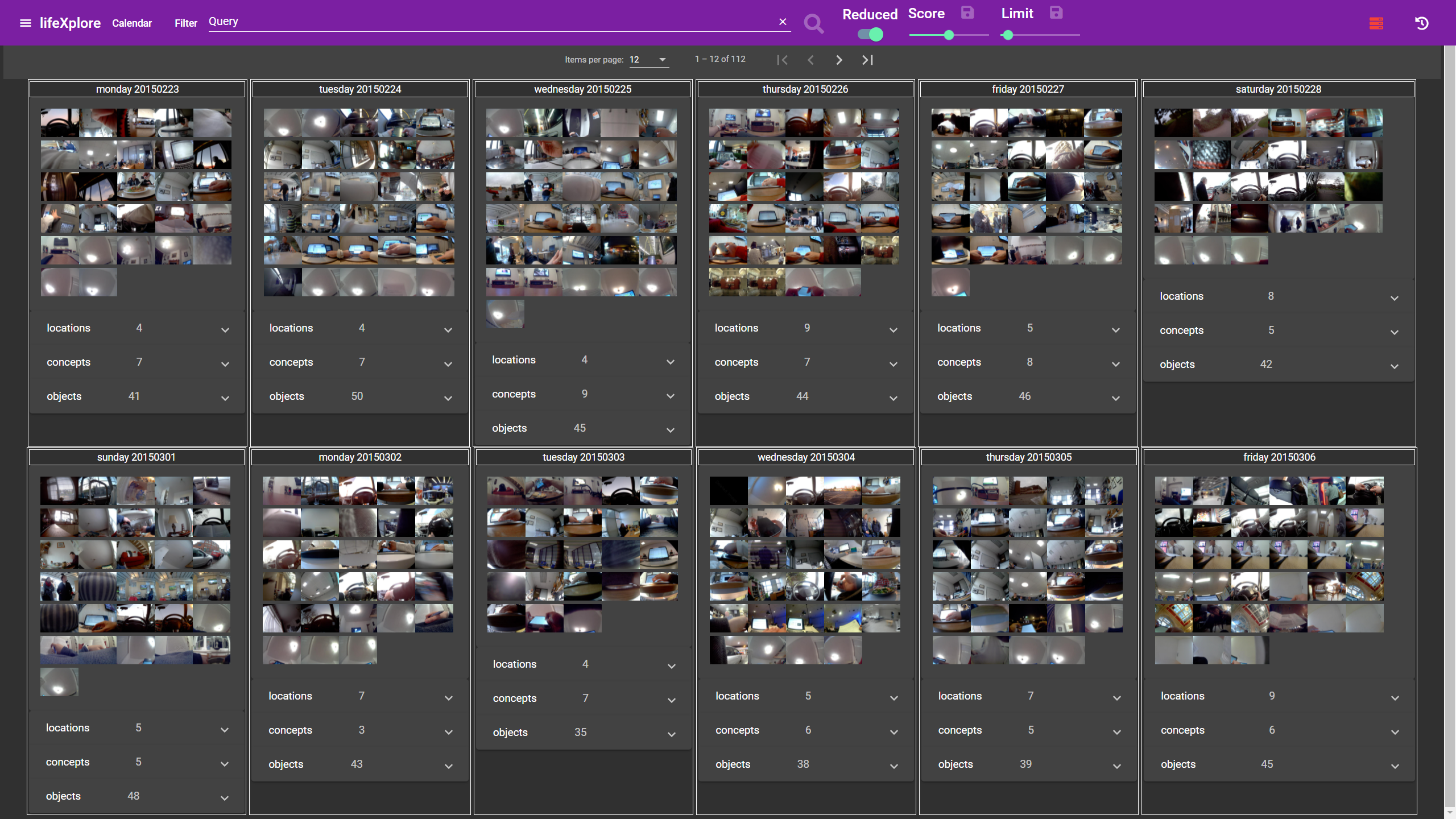}
      \caption{Calendar View with browseable day summaries}
      \label{fig:calendar_view}
    \end{subfigure}%
    \begin{subfigure}{.5\textwidth}
      \centering
      \includegraphics[width=\linewidth]{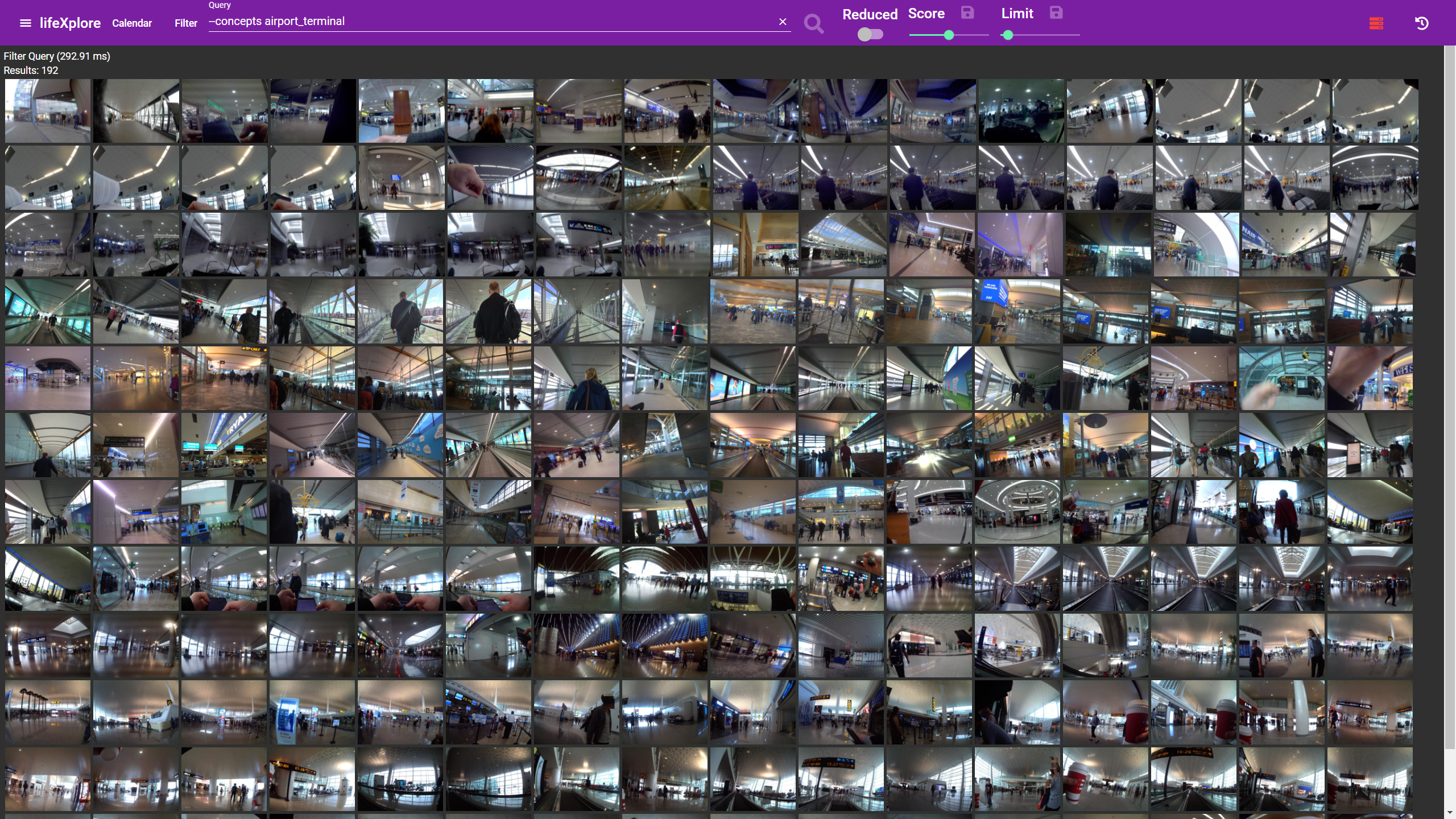}
      \caption{Filter View for combinable filter search}
      \label{fig:filter_view}
    \end{subfigure}

  \caption{lifeXplore's main interfaces}
  \label{fig:interface}
\end{figure*}

Albeit LSC2020 together with its hosting conference, the ACM ICMR'20, have been held virtually, the workshop was found  to be the most watched session of the entire conference, which highlights the growing public interest in this topic. Moreover, the increased participation of fourteen teams~\cite{DBLP:conf/mir/TranNB0G20,DBLP:conf/mir/MejzlikVKSL20,DBLP:conf/mir/HellerPGSS20,DBLP:conf/mir/KovalcikSSL20,DBLP:conf/mir/KhanLP0ZRKW20,DBLP:conf/mir/LeibetsederS20a,DBLP:conf/mir/Duane0G20,DBLP:conf/mir/LeNTNNZHG20,DBLP:conf/mir/TranNTTNNLTLNDV20,DBLP:conf/mir/RossettoBARPB20,DBLP:conf/mir/LiZMLM20,DBLP:conf/mir/ChuCYHC20,DBLP:conf/mir/Mai-NguyenPVTDZ20,DBLP:conf/mir/AlateeqRG20} marks an unprecedented rise in interested research parties. This is also reflected in the overall performance of the proposed systems: while the winning teams Mysc{\'{e}}al~\cite{DBLP:conf/mir/TranNB0G20} and SOMHunter~\cite{DBLP:conf/mir/MejzlikVKSL20} solved all 21 given tasks, on average the other systems managed to solve more than 14 tasks. After performing very poorly at LSC2019, we decided to refactor our continuously developed participating system, lifeXplore~\cite{munzer2018lifexplore,leibetseder2019lifexplore,DBLP:conf/mir/LeibetsederS20a}, which previously was based on diveXplore~\cite{munzer2018lifexplore,schoeffmannAutopiloting2019,leibetseder2020divexplore} -- an interactive video browsing and retrieval tool designed for the VBS challenge. In emphasizing simplicity as our main development goal, our system showed promising improvements by ranking 6th place on the leaderboard and we managed to solve 18 tasks. Thus, we feel confident to move in the right direction and will continue to further develop this system for LSC2021.

The lifeXplore system currently offers adjustable day summaries in a calendar-type view as well as combinable filters via autocompleting text input. Our contributions for the LSC2021 lifeXplore variant are the following

\begin{description}
    \item[Temporal Queries:] We add the functionality to combine separate queries in a temporal context, cf. Section~\ref{temporal_queries}.
    \item[Advanced Browsing:] We greatly improve the calendar view and add more advanced browsing capabilities, cf. Section~\ref{advanced_browsing}.
    \item[Usability Improvements:] We make the combinable filter view more user friendly, cf. Section~\ref{usability}.
\end{description}

In the following we describe the system in more detail in Section~\ref{lifexplore_description}, propose expert as well as novice strategies to solve a past LSC task in Section~\ref{solving_lsc} and highlight improvements to the previous system in Section~\ref{lifexplore_improvements}.

\section{The lifeXplore system}
\label{lifexplore_description}

The lifeXplore system is web-technologies based: it comprises a MongoDB\footnote{\url{https://www.mongodb.com/}} database that is searched via a Node.js\footnote{\url{https://nodejs.org/}} server, while being controlled by a user interface implemented in Angular\footnote{\url{https://angular.io/}}. Figure~\ref{fig:interface} shows the system's two main interfaces, \textit{Calendar View} and \textit{Filter View}. The former offers more freedom for data exploration and the latter is targeted at explicit search and filtering. Both of them, however, utilize the following analytical meta data:

\begin{description}
    \item[Time:] Simple dates are included but also weekday names and specifically defined timeframes such as morning, noon, afternoon etc.
    \item[Location:] Locations are based on geographic coordinates (latitude/longitude) but also metadata-based semantic names such as 'Home', 'Work' or 'The Helix'.
    \item[Deep Visual Features:] There are a variety of visual features including Concepts (e.g. 'airplane\_cabin' or 'staircase'), Attributes (e.g 'wet' or 'dry'), Objects (e.g. 'person' or 'bottle').
\end{description}

The subsequent sections describe the system's main views together with other useful components for solving tasks.

\subsection{Calendar View}

The Calendar View lists all contained days of the dataset in a browseable manner. Users are presented with picture summaries of each date and also its most prominent locations, concepts and objects in terms of occurrence frequency. As Figure~\ref{fig:calendar_view} shows, these day summaries are navigated with a paginator that also allows for choosing the number of days to display. This provides a user with the possibility of inspecting fewer dates with greater detail. Additionally, the displayed days can be filtered by weekday name in order to facilitate a quicker browsing experience, while omitting potentially irrelevant data.

\begin{figure}[htbp!]
    \centering
    \includegraphics[width=\linewidth]{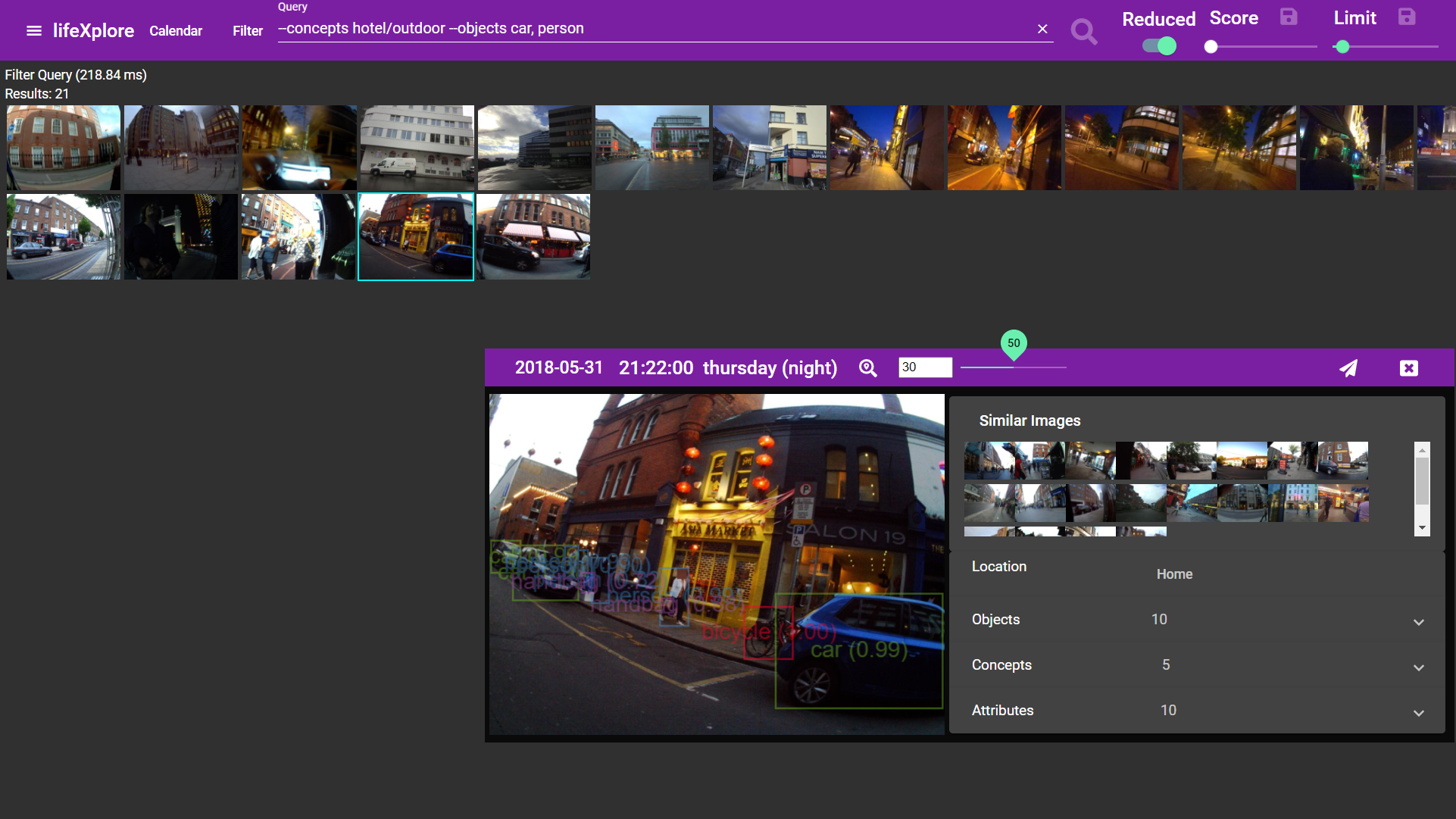}
    \caption{Combinable filters - search for concept 'hotel/outdoor' in combination with objects 'car' and 'person'}
    \label{fig:filtered_result}
\end{figure}


\subsection{Filter View}

The Filter View, depicted in Figure~\ref{fig:filter_view}, can be considered the heart of the system. Here, users can search and combine all of the aforementioned metadata in order to greatly reduce the number of retrieved results. This is accomplished via a simple textual input, which behaves similarly to an argument parser of a typical command line application: a user types in a keyword such as '\texttt{-{}-objects}' or simply '\texttt{-o}' followed by a comma separated list of search terms such as '\texttt{apple,banana}' with the intention of retrieving images that contain both of these objects. Optionally, dependent on the utilized keyword, these terms can also contain individual confidence scores, which offer great flexibility in defining searches: for instance, '\texttt{-o apple(0.9),banana}' would issue a search for apples that are detected with at least a confidence of 90\% and bananas with the globally defined system confidence, which can be adjusted with the interface's 'Score' slider: in Figure~\ref{fig:filter_view} this is set to a very low setting of approximately 10\%. This strategy is applicable for Concepts, Attributes and Objects, albeit confidence scores can only be applied to metadata containing this statistic. Other filters can be used in a similar manner, however, they always are executed according to the most intuitive interpretation: for example, when searching for images taken on weekends a user can simply type '\texttt{-{}-weekdays saturday,sunday}' and the system returns results for Saturdays or Sundays instead of creating an empty list after a futile attempt to find days that qualify for both of those week days. While choosing to combine several of these search strategies, users can additionally limit the number of returned results using the 'Limit' slider as well as filter out similar images with the 'Reduced' switch. Finally, Figure~\ref{fig:filter_view} shows an example of combining multiple filters together with limiting the results in terms of quantity and similarity: '\texttt{-{}-concepts hotel/outdoor --objects car,person}'.

\begin{figure}[htbp!]
    \centering
    \includegraphics[width=\linewidth]{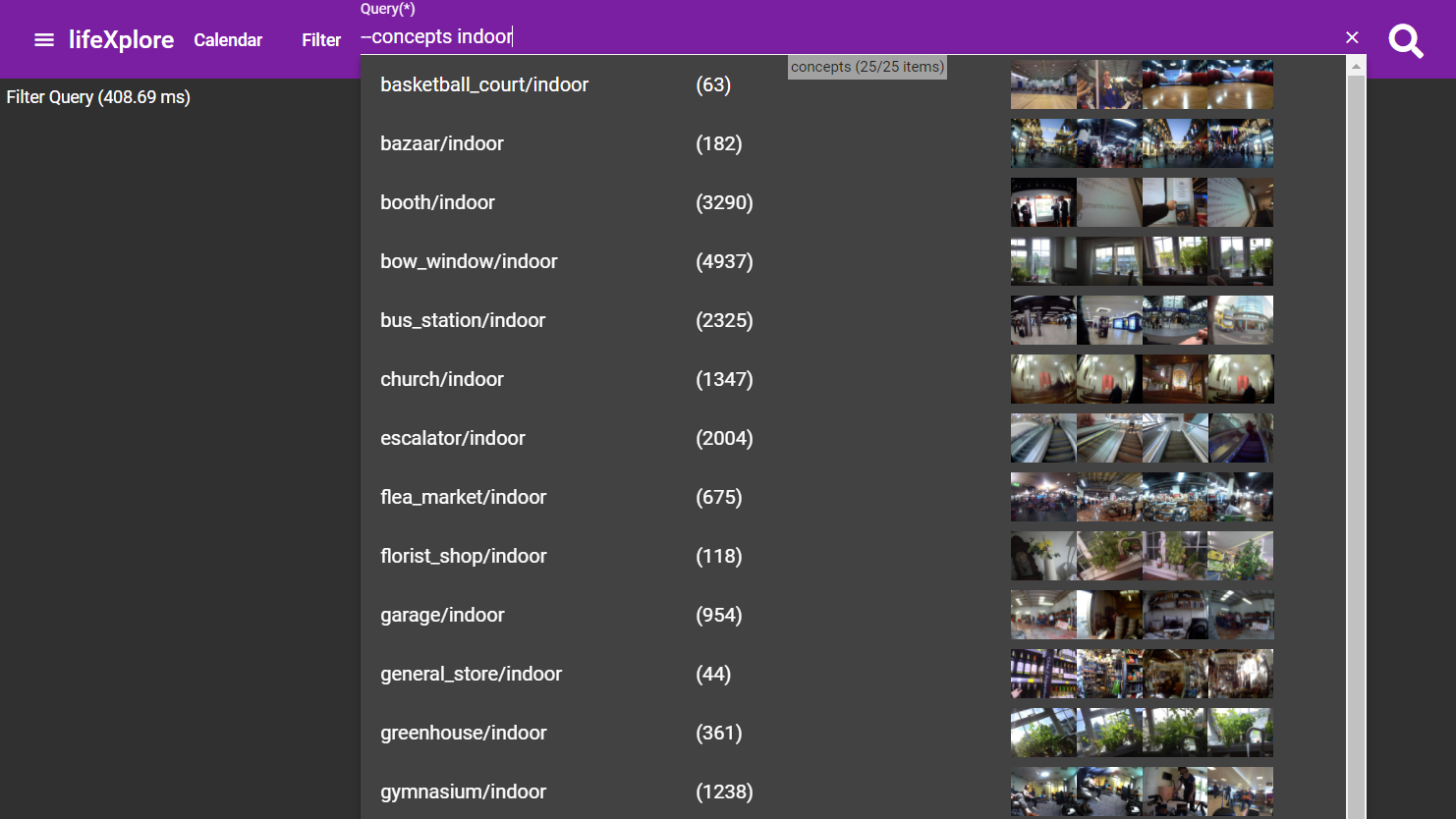}
    \caption{Autocomplete feature providing dynamically updated suggestions and information during user input}
    \label{fig:autocomplete}
\end{figure}

\subsection{Autocompleting Input}

Since it is impossible for a user to know all of the deep visual classes and other metadata details that are stored in the database, the system offers a smart autocompletion feature that is dynamically updated according a user's input. For instance, as shown in Figure~\ref{fig:autocomplete}, it is triggered while a user is trying to find concepts that include the word '\texttt{indoor}', in which case a dropdown menu provides the user with 80 selectable suggestions, statistic about how many instances of the item is contained in the database as well as several example images that help the user better understand the system's conception about individual features. Apart from being helpful with individual Concepts, Objects or Attributes, autocompletion simply can be triggered to show all keywords the system knows as well as their aliases by simply typing '\texttt{-}' or '\texttt{-{}-}'. Moreover, autocompletion also conveys the user with the meaning of not easily interpretable concepts such as listing the actual timeframe as well as number of images for '\texttt{-{}-timename afternoon}', i.e. a window between 13:00 17:00 comprising 48 420 images. Lastly, this feature also helps with values for more complex keywords like '\texttt{--location}', which certainly could be utilized via entering latitude and longitude (as can be auto-inserted using the Image View, cf. Section~\ref{sec:image_view}) but in practice is much more intuitively used by focusing on the available named locations, as are suggested by the system.

\begin{figure}[htbp!]
    \centering
    \includegraphics[width=\linewidth]{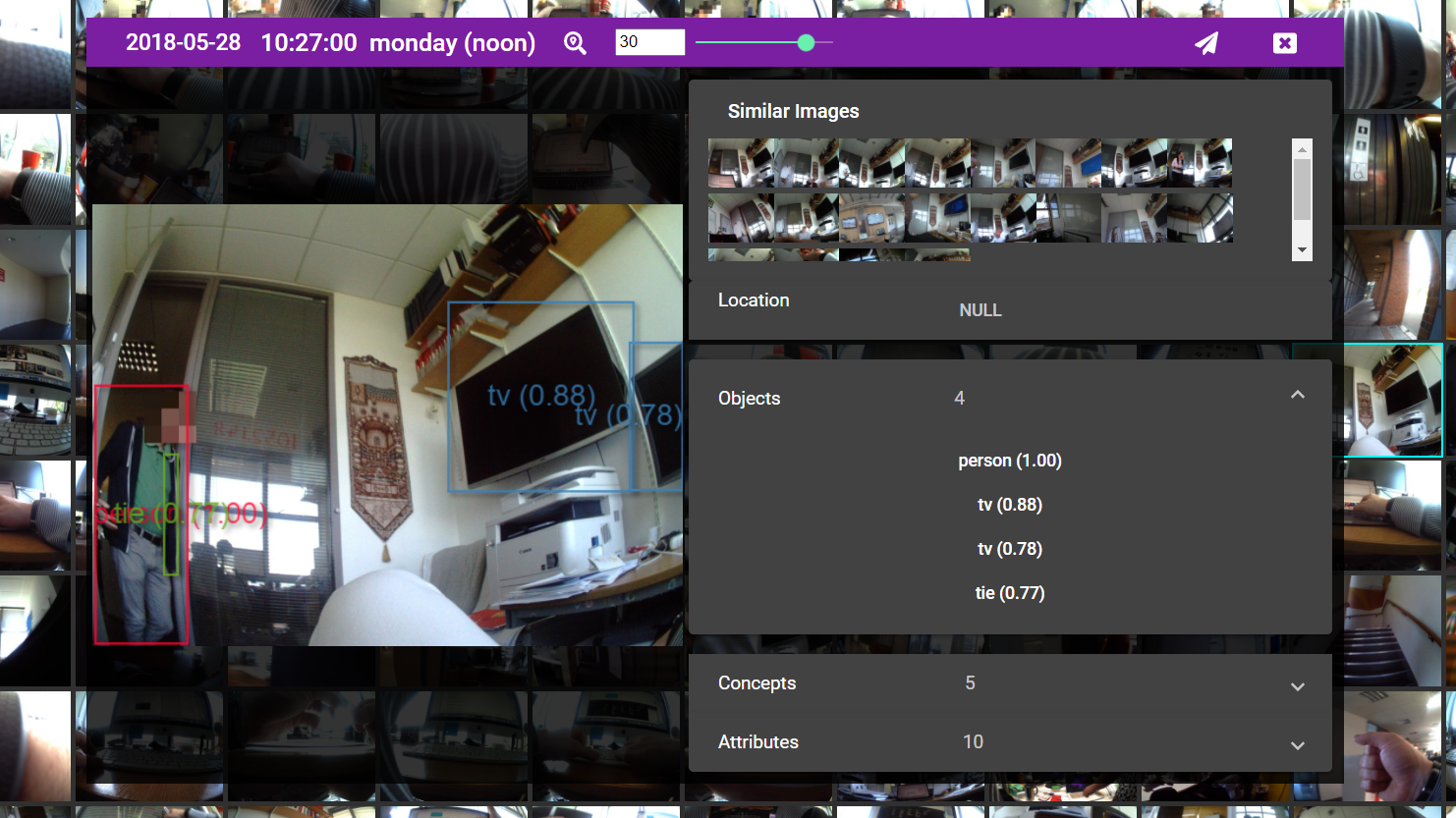}
    \caption{Image view with similar images, statistics, object bounding boxes and built-in links for follow-up searches}
    \label{fig:image_view}
\end{figure}

\subsection{Image View}
\label{sec:image_view}

The Image View -- an overlay window for detailed inspection of results (shown in Figures~\ref{fig:image_view} and~\ref{fig:filtered_result}) -- is triggered by clicking on any thumbnail that is part of the retrieved result items. Once opened, users can even navigate the entire grid of results using the arrow keys on their keyboard, while the Image View is updated according to the currently selected item (cyan-bordered image in Figure~\ref{fig:image_view}). The Image View overlay contains all necessary information about the selected image, such as date, time and -- if available -- named location. Additionally to an enlarged image version, bounding boxes enclosing identified objects can be displayed with adjustable degrees of transparency (green slider). Furthermore, all identified deep visual features are listed in a series of expandable panels and, if available, a specific additional panel lists similar images to the opened one. Besides merely displaying all of this information, the view also provides convenient and even crucial functionality. For example, a submission can only be made in this view exclusively by clicking the icon resembling a paper plane. Most importantly, nearly every information is provided as a clickable link, which triggers a new query to be issued in the current browser window or, optionally, in a new one. Albeit the provided links mostly trigger basic searches depending on the currently displayed item, such as retrieving all images of its date, querying dates of similar images or searching for other images containing the same objects, these queries are automatically entered into the user input and, therefore, can be quickly refined further along the way. Finally, a very convenient alternative to working with deep visual features is the Image View's radius-based geographic location search functionality, which is especially useful when named locations are missing: clicking the magnifier icon issues a search for images taken within an adjustable circular area around the source image (30 kilometres in Figure~\ref{fig:image_view}). This can bring an advantage for searching items where geographic locations vary strongly, e.g. traveling days.

\begin{figure}[htbp!]
    \centering
    \includegraphics[width=\linewidth]{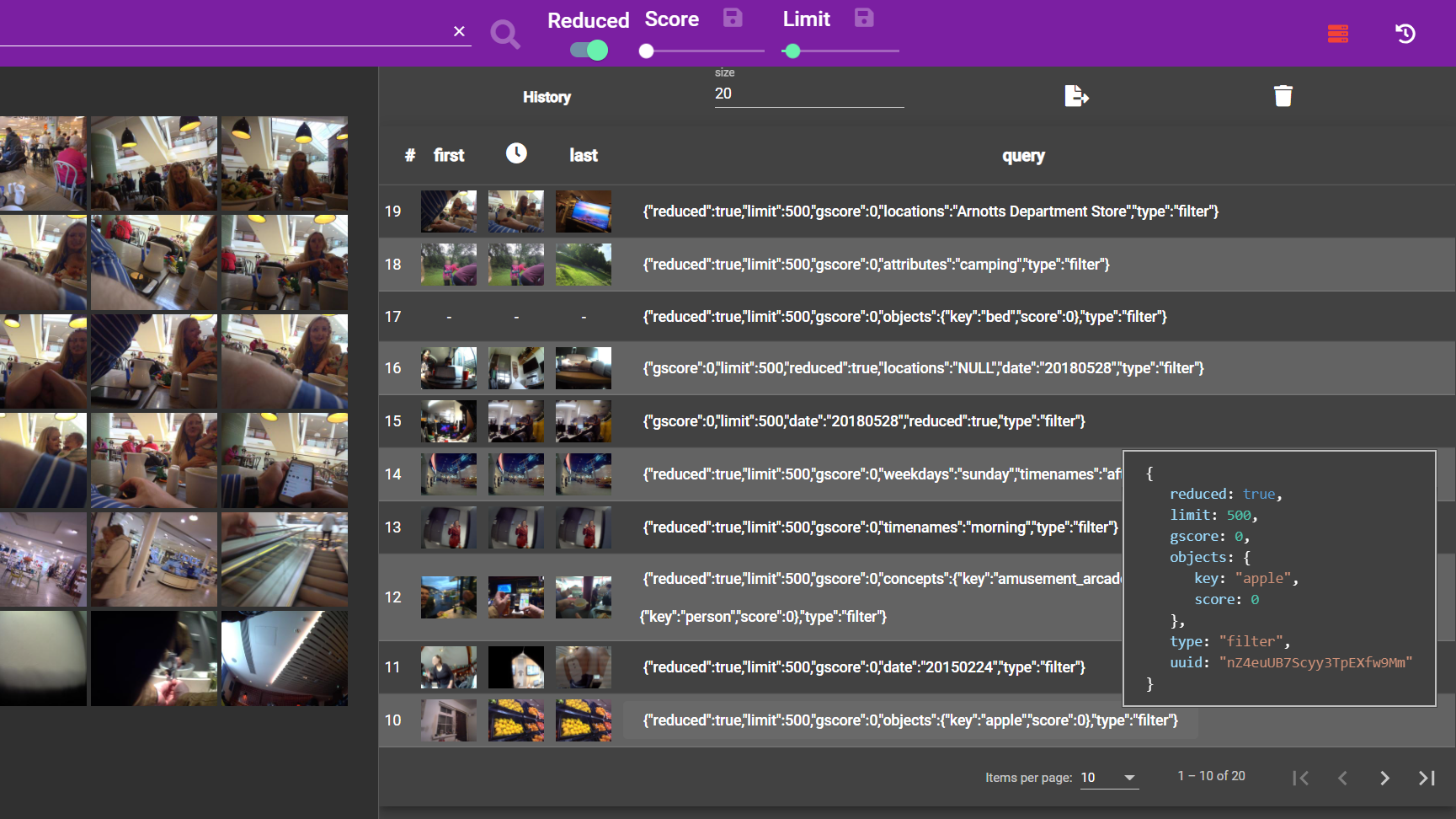}
    \caption{History listing all searches as well as associated opened images according to first, last and longest viewed}
    \label{fig:query_history}
\end{figure}

\subsection{Query History}

As depicted in Figure~\ref{fig:query_history}, the Query History component is provided as a convenient sidebar that can be opened and closed at any time during a session. It records every query issued together with the first, last and longest viewed images that the user inspected using the Image View. Any recorded query can be re-issued by clicking on its associated JSON record holding all set parameters for querying the database server. Furthermore, by clicking on any of the images, the query is not only run again but the system automatically scrolls to that image once the results are loaded. To ensure that the Query History does not create duplicate records, contained entries carry a unique ID -- running a recorded query again only moves it to the top of the list marking it as the most recent query. Lastly, there is a configurable upper limit of history items and they are stored in the local browser storage. This prevents continuously filling up the local storage as well as making the application impervious to closing the app during a session. Nevertheless, a user can of course explicitly trigger a deletion of the history in order to start a new session.

\begin{figure*}[htbp!]

    \centering
    \begin{subfigure}{0.33\textwidth}
        \centering
        \begin{tikzpicture}[spy using outlines={chamfered rectangle, yellow, magnification=3, width=2.4cm, height=1cm, connect spies}]
            \node {\pgfimage[width=0.98\linewidth]{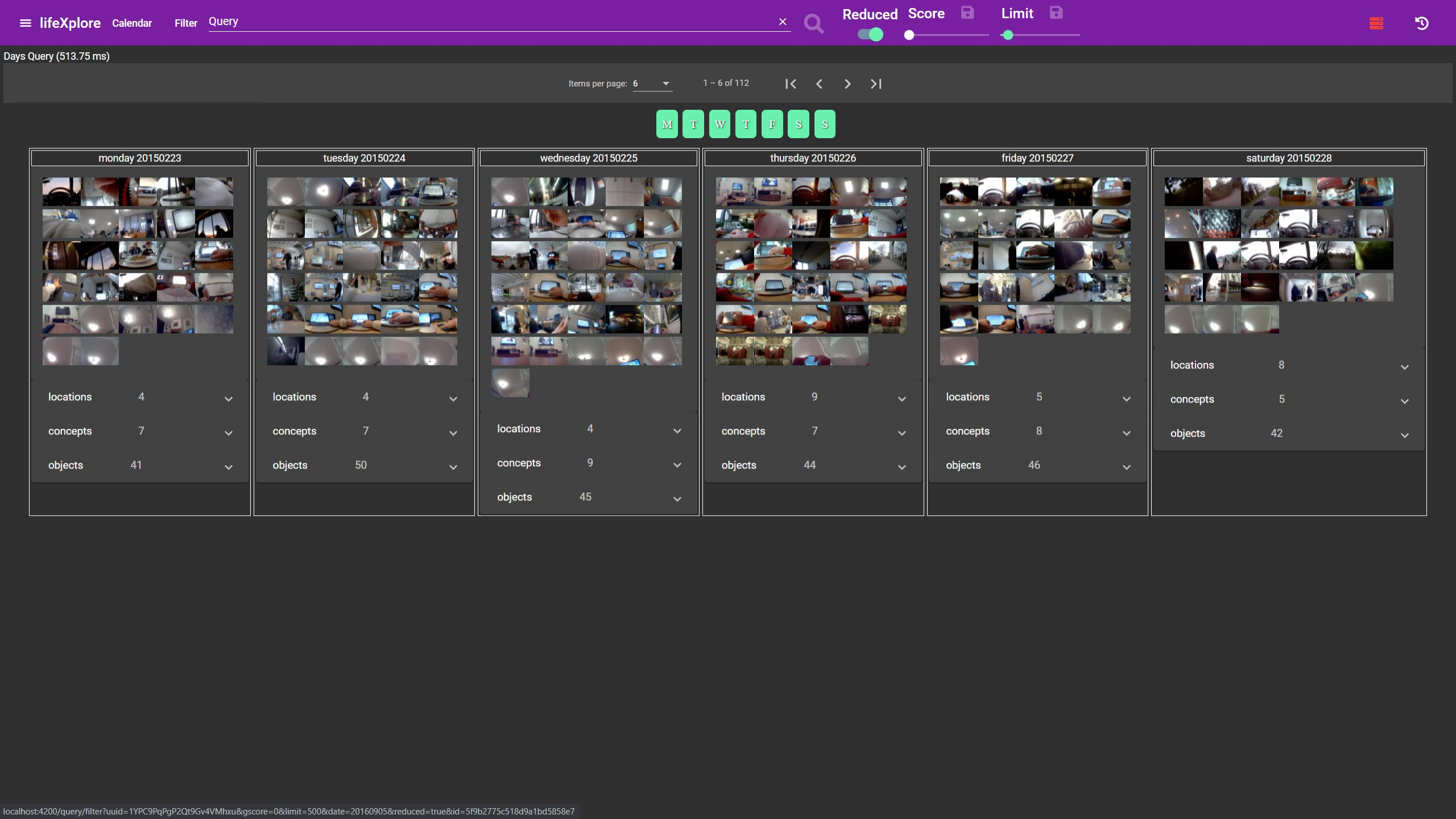}};
            \spy on (0.07, 1.1) in node [right] at (0.3, -1.0);
        \end{tikzpicture}
        \caption{N1 -- open calendar view}
        \label{fig:novice_1}
    \end{subfigure}%
    \hfill
    \begin{subfigure}{.33\textwidth}
        \centering
        \begin{tikzpicture}[spy using outlines={chamfered rectangle, yellow, connect spies}]
            \node {\pgfimage[width=0.98\linewidth]{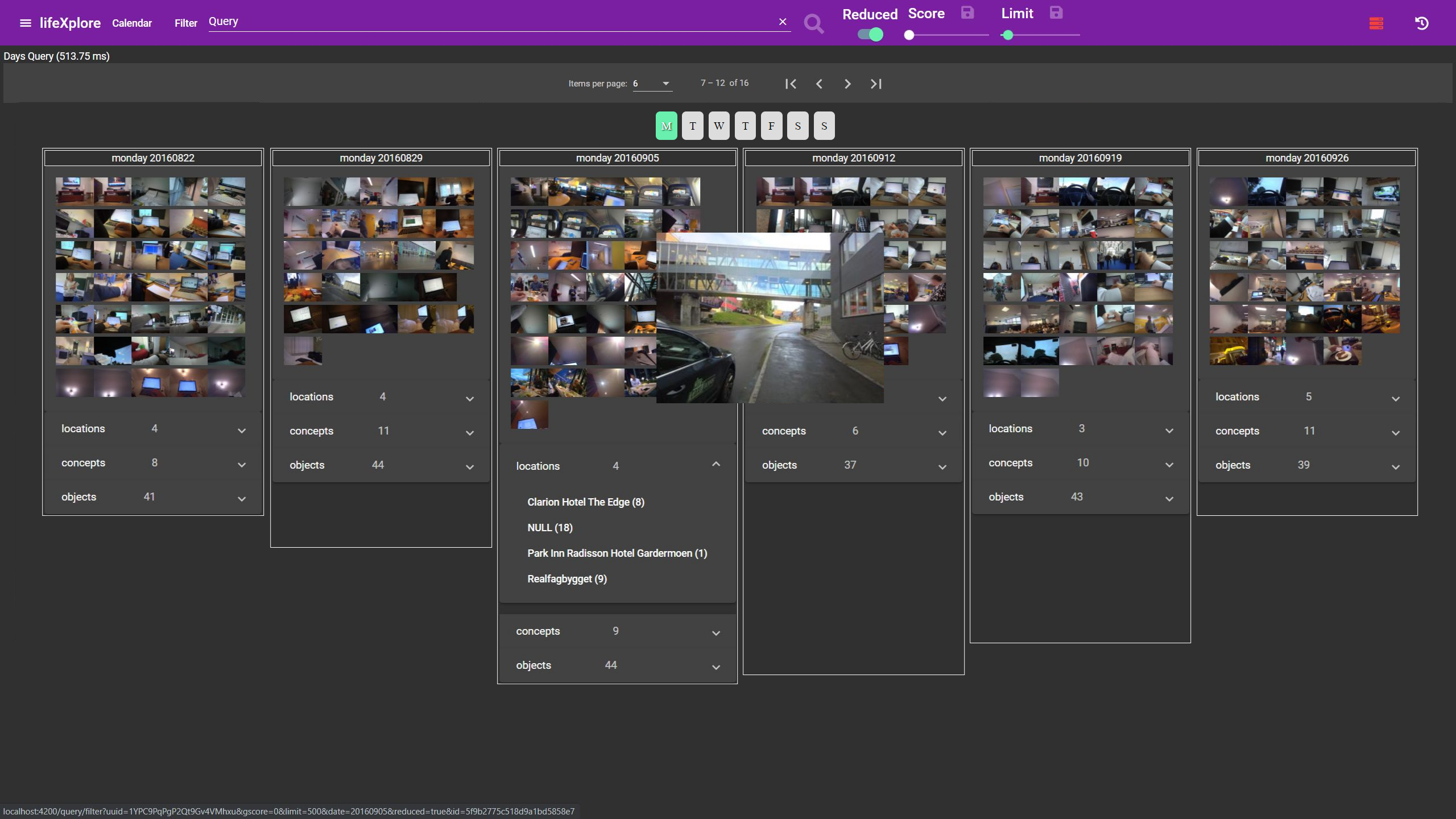}};
            \spy[magnification=3, width=2.4cm, height=1cm] on (0.07, 1.1) in node [right] at (0.3, -1.0);
            \spy[magnification=2, width=2.2cm, height=1.7cm] on (0.07, 0.4) in node [left] at (-0.6, -0.7);
        \end{tikzpicture}
      \caption{N2 -- browse all 16 'Monday' day summaries}
      \label{fig:novice_2}
    \end{subfigure}
    \hfill
    \begin{subfigure}{.33\textwidth}
      \centering
        \begin{tikzpicture}[spy using outlines={chamfered rectangle, yellow, magnification=2, width=2.3cm, height=1.9cm, connect spies}]
            \node {\pgfimage[width=0.98\linewidth]{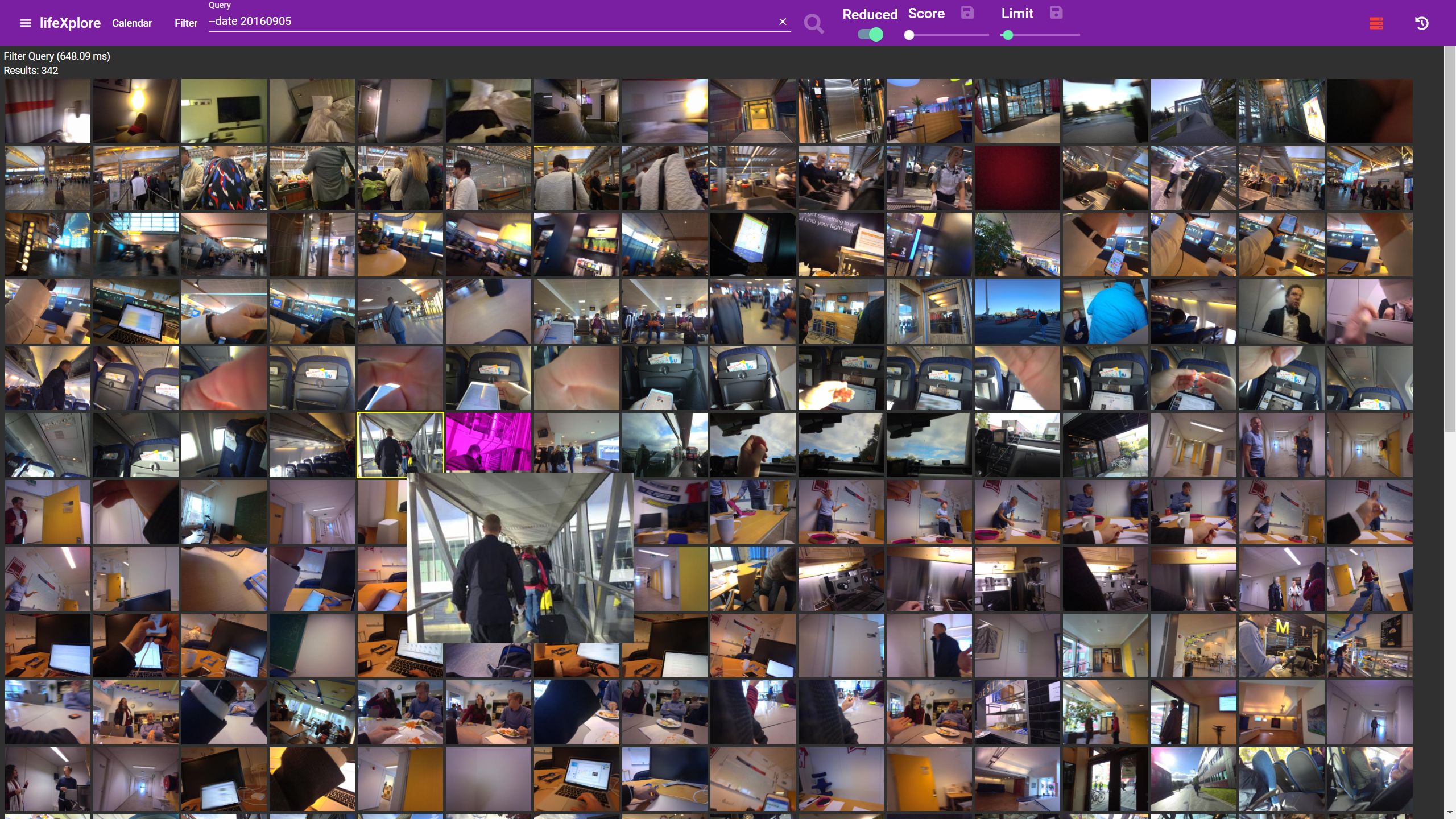}};
            \spy on (-0.91, -0.47) in node [right] at (0.4, 0.40);
        \end{tikzpicture}
      \caption{N3 -- day 2016/09/05, similar images omitted}
      \label{fig:novice_3}
    \end{subfigure}

    \vskip\baselineskip

    \begin{subfigure}{.33\textwidth}
      \centering
        \begin{tikzpicture}[spy using outlines={chamfered rectangle, yellow, magnification=3, width=2.4cm, height=0.6cm, connect spies}]
            \node {\pgfimage[width=0.98\linewidth]{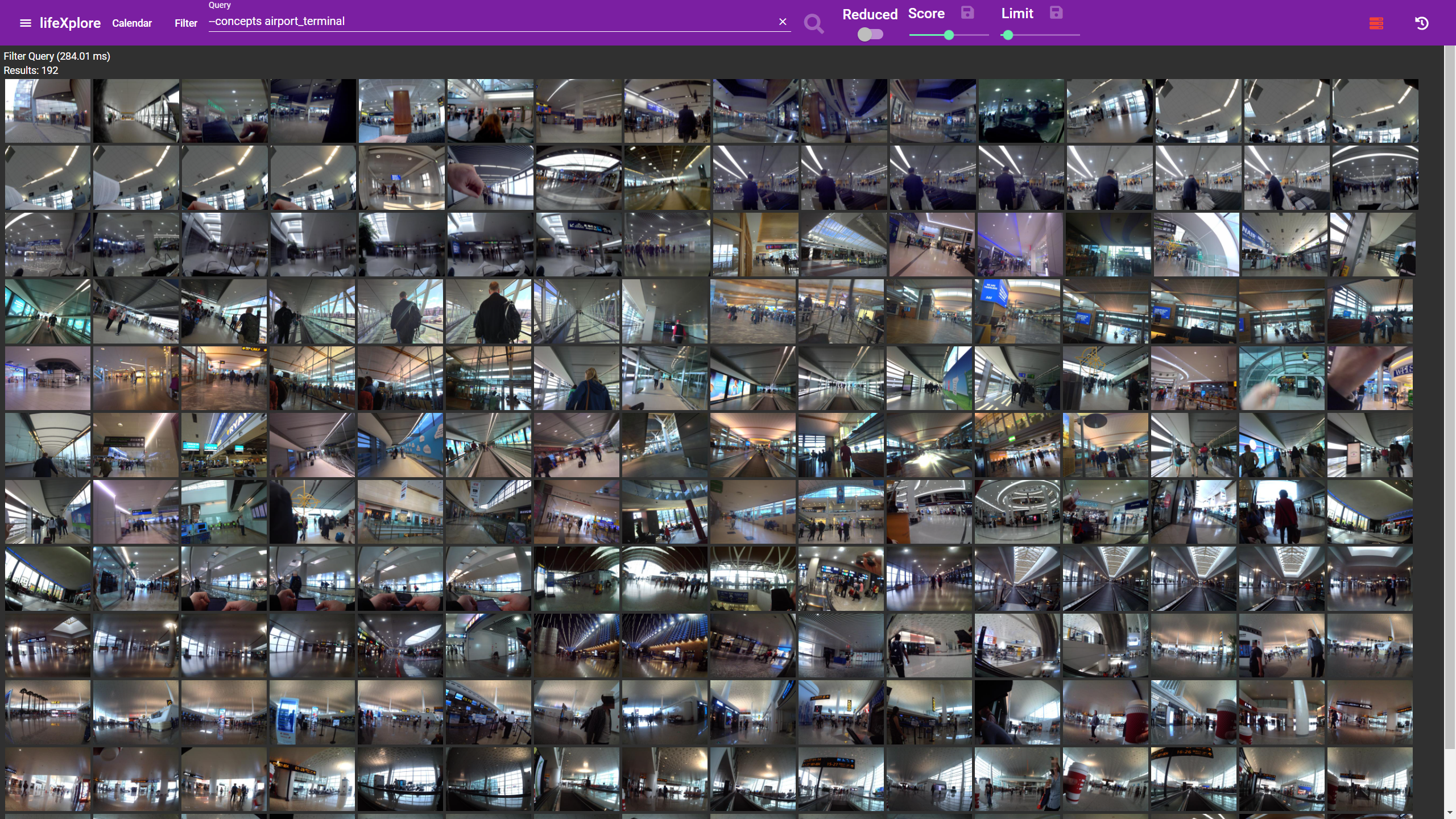}};
            \spy on (-1.85, 1.53) in node [right] at (0.35, 1.05);
        \end{tikzpicture}
      \caption{E1 -- search for concept 'airport\_terminal'}
      \label{fig:expert_1}
    \end{subfigure}
    \hfill
    \begin{subfigure}{.33\textwidth}
      \centering
        \begin{tikzpicture}[spy using outlines={chamfered rectangle, yellow, magnification=3, width=3.5cm, height=0.6cm, connect spies}]
            \node {\pgfimage[width=0.98\linewidth]{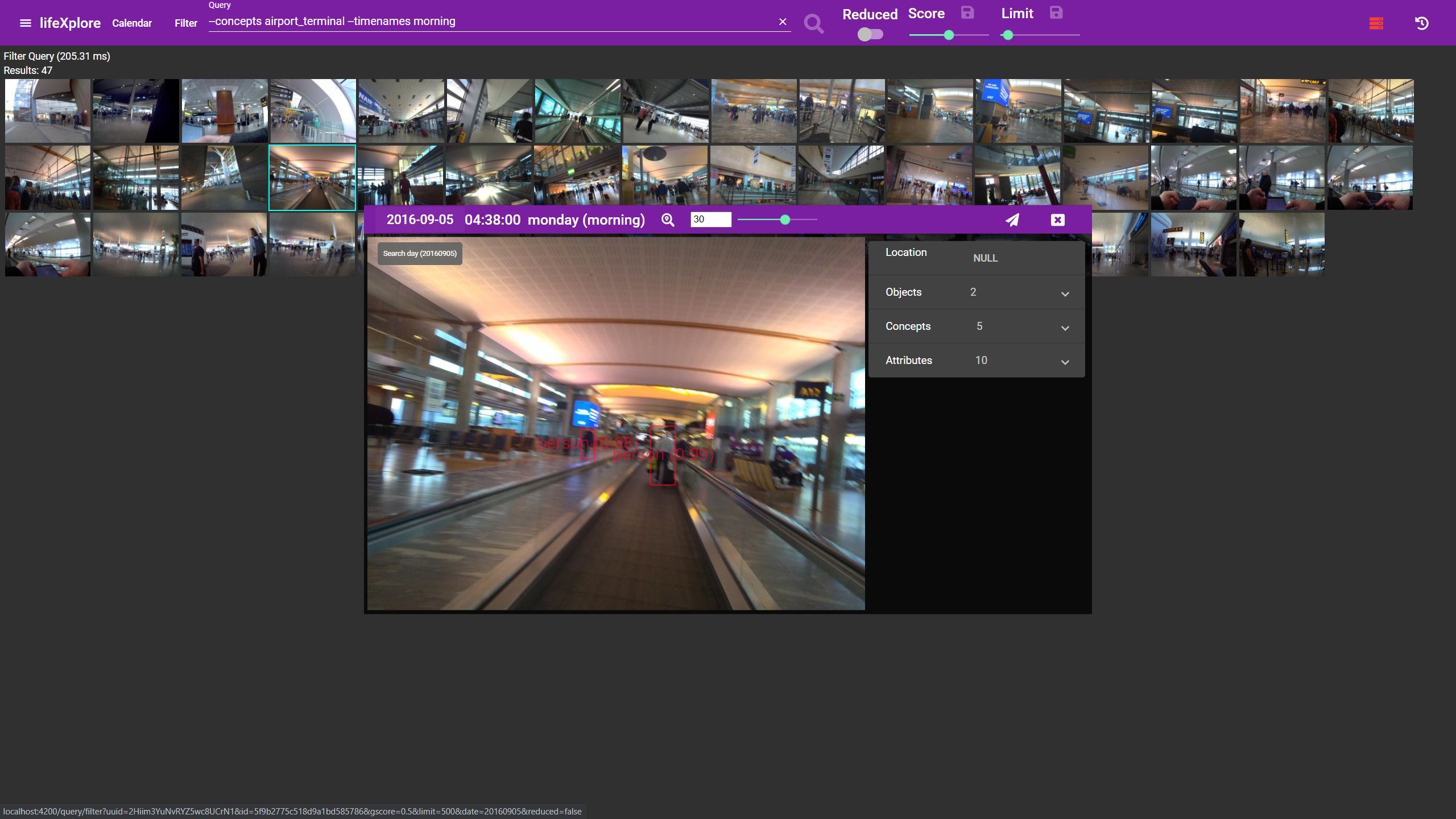}};
            \spy on (-1.63, 1.53) in node [right] at (-0.7, 1.05);
        \end{tikzpicture}
      \caption{E2 -- add named time 'morning'}
      \label{fig:expert_2}
    \end{subfigure}
    \hfill
    \begin{subfigure}{.33\textwidth}
      \centering
        \begin{tikzpicture}[spy using outlines={chamfered rectangle, yellow, magnification=2, width=2.8cm, height=2.4cm, connect spies}]
            \node {\pgfimage[width=0.98\linewidth]{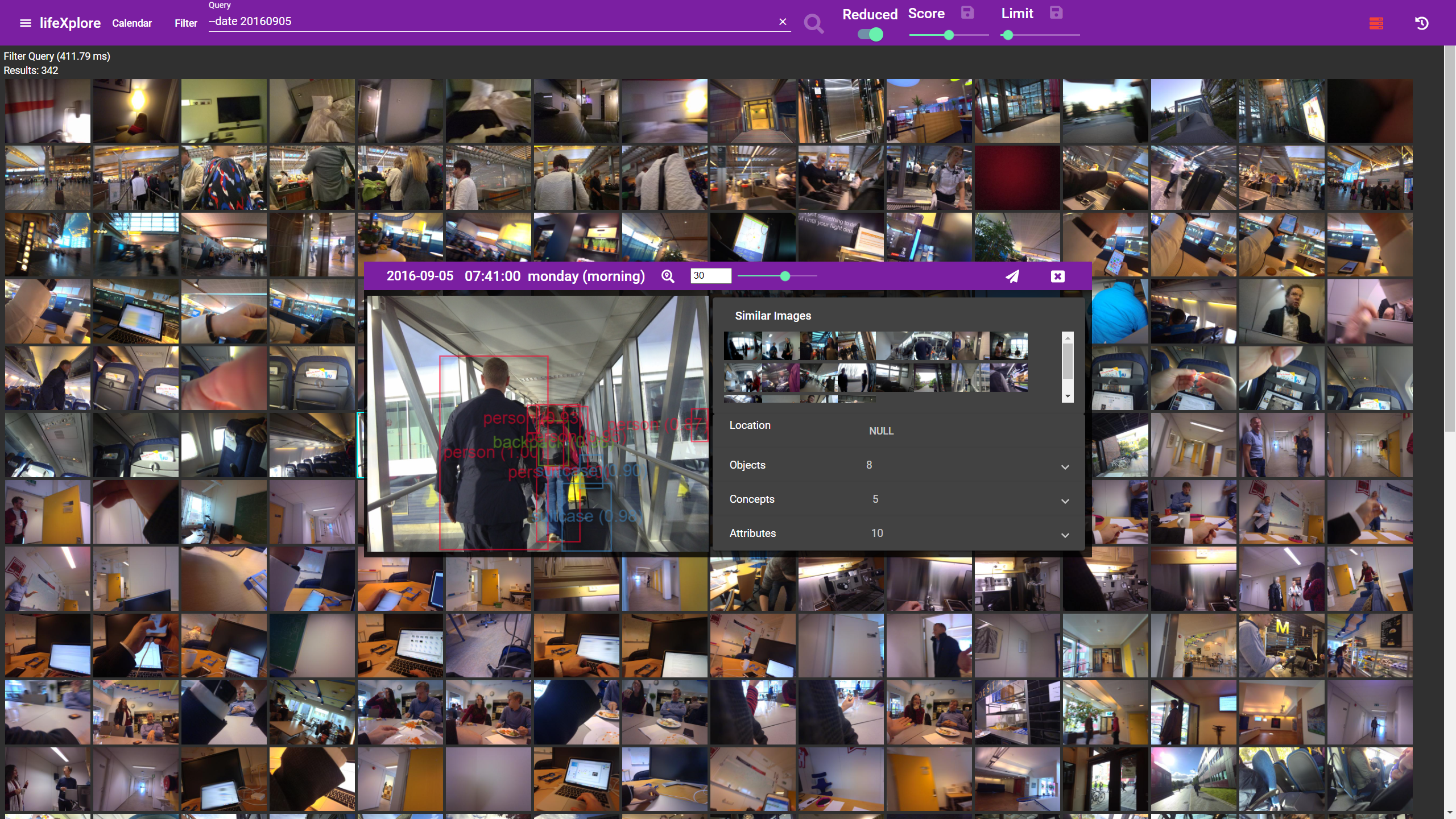}};
            \spy on (-0.75, 0.0) in node [right] at (0, 0.1);
        \end{tikzpicture}
      \caption{E3 -- day 2016/09/05, result in Image View}
      \label{fig:expert_3}
    \end{subfigure}

  \caption{Solving an LSC task as a Novice (N) and Expert (E)}
  \label{fig:search_strategy}
\end{figure*}

\section{Solving a Lifelogging Task}
\label{solving_lsc}

In this section, we demonstrate how the system can be utilized to solve a past lifelogging task as a novice as well as expert user, also giving a visual comparison in Figure~\ref{fig:search_strategy}. The task has a total duration of 3 minutes (180 seconds) and it gradually is made easier over time and new information is made available to the participants at certain timestamps ('t' in seconds):

\begin{itemize}
    \item[\bf t=0] Walking through an airbridge after a flight
    \item[\bf t=30] in the early morning
    \item[\bf t=60] after a flight of about two hours.
    \item[\bf t=90] After the airport, I immediately got a taxi to a meeting.
    \item[\bf t=120] I think it was a cloudy day on a Monday.
    \item[\bf t=150] I was in Tromso in Norway.
\end{itemize}

\subsection{Novice Strategies}

As a novice might not be very familiar with the filtering mechanics of the system, hence, potentially starts out using the Calendar View, as depicted in Figure~\ref{fig:novice_1} depicts. Choosing to reduce the number of displayed days to 6 helps the user for a better focusing on the summary views. Here it is important to regard days with noticeable more locations, as the text early on already hints that a flight has taken place. Additionally, the novice is looking out for pictures that indicate a flight cabin or an airport terminal, which is also supported by expanding each day's most frequently seen concepts menu. Continuing this strategy, as more information becomes available, the user pays special attention to car rides and meeting rooms that follow flight scenarios. The most crucial information is revealed at $t=120$, after which the novice is able to restrict the displayed summaries to only contain Mondays, which greatly reduces the number of days to browse down to 16. Luckily the day in question is already found in the second set of days (cf. Figure~\ref{fig:novice_2}) and the user is able to open this day in the Filter View (cf. Figure~\ref{fig:novice_2}). Finally, with this view's 'Reduced' option the user can quickly identify the scene in the result grid. Albeit less effective for specific concepts, such a strategy enriches the familiar traditional usage of a digital calendar with visual aids and semantic concepts. 

\subsection{Expert Strategies}

Experts typically start out on the combinable filter view, issuing a Concept-query for 'airport\_terminal' (cf. Figure~\ref{fig:expert_1}), as automatically suggest by the system after entering a few letters. The system returns many potential results and only at $t=30$, the filter can be refined to include the named time 'morning', which greatly reduces the results to 47 images (cf. Figure~\ref{fig:expert_2}). At this point, all that is left to do is finding the right day, which may prove challenging, yet, using the day links via the Image View for opening different days in new browser windows most likely is faster than reaching $t=120$, where the additional weekdays filter for 'Monday' can be added. Either way, the expert eventually reaches the targeted day and opens the image most likely to be within the described scene for submission via the Image View (cf. Figure~\ref{fig:expert_3}). Overall, this strategy most probably is faster but both, expert and novice approaches are very likely to succeed in finding the scene in question, at least with the current dataset including a manageable number of Mondays.

\section{Improvements}
\label{lifexplore_improvements}

For LSC2021, we implement a number of improvements, which are described in below sections. 

\subsection{Temporal Queries}
\label{temporal_queries}

In light of the gradually updating task progression often describing scenes in a temporal manner, we leverage the already inherent temporal dependencies of images as well as metadata for further improving filter queries. Users are able to set multiple queries in a temporal context to each other, which enables searches to be refined in the same manner as the given tasks. A particular challenge of adding this feature is to adjust the user input to incorporate such dependencies, also making it easy to expand and reduce lists of temporally dependent queries. The most simple solution when regarding the quite complex smart user input is allowing users to dynamically add, remove and reorder such inputs, essentially providing an interface for multi-input queries. This feature also entails refactoring the Query History component as to being able to store and re-issue multiple temporally dependent queries.

\subsection{Advanced Calendar View}
\label{advanced_browsing}

As the Calendar View is quite basic, it is improved by including more sorting options such as most frequent concept, object or location. Albeit less powerful than the Filter View, this view still has the advantage of separating the content of individual days, which prevents users from confusing dates with similar-looking images. In addition, the number of images per day summary is made adjustable in order to allow for exploration at different levels of detail.

\subsection{Usability Improvements}
\label{usability}

Smaller usability improvements are integrated into the Filter View in order to allow for a better novice experience. This includes hiding buttons that might be confusing for a novice as well as initializing the system with a recommended default configuration that supports inexperienced users well. Finally, the result grid is improved by adding different sorting options such as date, confidence score or number of found objects.

\section{Conclusions and Future Work} \label{sec: conclusions}

With the intent on participating in LSC2021, we present our lifeXplore system, which combines exploring a lifelog database chronologically via calendar-inspired day summaries as well as in a content-based manner via applying combinable filters. For future work, we plan on incorporating mechanisms that address the increasing amount of lifelog data employed in the LSC competition. In particular, the scalability of our backend database will be subject to thorough investigations and improvements.

\begin{acks}
This work was funded by the FWF Austrian Science Fund under grant P 32010-N38.
\end{acks}

\balance

%
\bibliographystyle{ACM-Reference-Format}
\bibliography{bibliography}

\end{document}